\documentstyle[emulateapj,epsf]{article}

\def\agt{\mathrel{\raise.3ex\hbox{$>$}\mkern-14mu\lower0.6ex\hbox{$\sim$}}}
\def\alt{\mathrel{\raise.3ex\hbox{$<$}\mkern-14mu\lower0.6ex\hbox{$\sim$}}}

\newcommand{\beq}{\begin{equation}}
\newcommand{\eeq}{\end{equation}}
\newcommand{\beqn}{\begin{eqnarray}}
\newcommand{\eeqn}{\end{eqnarray}}
\newcommand{\pa}{\partial}

\begin{document}

\title{Stability of rigidly rotating supermassive 
  stars against gravitational collapse}

\author{Masaru Shibata \altaffilmark{1}, Haruki Uchida \altaffilmark{1}, and 
Yu-ichiro Sekiguchi \altaffilmark{2}
}

\affil{\altaffilmark{1} 
Yukawa Institute for Theoretical Physics, Kyoto University, 
Kyoto 606-8502, Japan}

\affil{\altaffilmark{2} 
Department of Physics, Toho University, Funabashi, Chiba 274-8510, Japan}

\begin{abstract}

We revisit secular stability against quasi-radial collapse for rigidly
rotating supermassive stars (SMSs) in general relativity. We suppose
that the SMSs are in a nuclear-burning phase and can be modeled by
polytropic equations of state with the polytropic index $n_p$ slightly
smaller than $3$. The stability is determined in terms of the
turning-point method.  We find a fitting formula of the stability
condition for the plausible range of $n_p$ ($2.95 \alt n_p \alt 3$)
for SMSs. This condition reconfirms that, while non-rotating SMSs with
mass $\sim 10^5M_\odot$--$10^6M_\odot$ may undergo a
general-relativistically induced quasi-radial collapse, rigidly
rotating SMSs with a ratio of rotational to gravitational potential
energy ($\beta$) of $\sim 10^{-2}$ are likely to be stable against
collapse unless they are able to accrete $\sim 5$ times more mass
during the (relatively brief) hydrogen-burning phase of their
evolution.  We discuss implications of our results.

\end{abstract}


\keywords{relativity -- hydrodynamics -- stars: rotation}


\section{Introduction}

A supermassive star (SMS) is a possible progenitor for the formation
of a seed of a supermassive black hole (SMBH). Recent star-formation
calculations in spherical symmetry (Hosokawa et al.,~2013; Umeda,
private communication) suggest that if a high mass-accretion rate with
$\agt 0.1M_\odot$/yrs is preserved in the period of nuclear-burning
phases $\sim 2 \times 10^6$\,yrs, a SMS with mass $\agt 2 \times
10^5M_\odot$ could be formed.  Such a high mass-accretion rate
requires primordial gas clouds with virial temperature $\agt
10^{4}$\,K.  There are several scenarios proposed to achieve this
condition such as Lyman-Werner radiation from nearby local star
formation region (Omukai, 2001; Dijkstra et al.,~2008) or shock
heating in the cold accretion flows in the forming first galaxies
(Dekel et al.~2009; Inayoshi \& Omukai,~2012). Subsequently, the SMS
may collapse to a seed of a SMBH of mass $\agt 10^5M_\odot$.  As the
mechanism of the collapse of the SMSs, general-relativistic radial
instability (Iben,~1963; Chandrasekhar,~1964; Zel'dovich \& Novikov, 1971)
is often referred.

The formation process of a SMBH after the collapse of a SMS should be
determined by the initial condition at which the instability sets in.
In reality, it is natural to consider that SMSs are rotating because
they are likely to be formed in a non-symmetric environment at a dense
core of the galactic center as indicated by recent numerical
simulations of the collapse of an atomic cooling halo in the early
Universe (e.g., Latif et al.~2013; Regan et al.~2014; Becerra et
al.~2015). These simulations have suggested that proto-stellar disks
initially formed in the central gas cloud could be gravitationally
unstable and fragment into several clumps, preventing the growth of
both mass and angular momentum of the central protostar.  However, the
clumps are likely to subsequently migrate inward and eventually fall
onto the central protostar, enhancing episodic accretion (Inayoshi \&
Haiman,~2014; Hosokawa et al.,~2015): A rotating SMS could be a likely
outcome.

This implies that for realistic exploration of the collapse of SMSs to
a SMBH, we have to derive the stability condition for {\em rotating}
SMSs. This is in particular the case in this context because the SMSs
are very massive, and hence they are supported dominantly by the
radiation pressure, resulting in the adiabatic index, $\Gamma$, close
to $4/3$.

The condition for the stability of rotating SMSs was first analyzed by
Fowler~(1966) in his post-Newtonian analysis (see also chapter~14 of
Tassoul, 1978 for a review). He showed that the rotation plays a
significant role for stabilizing the radiation-supported SMSs against
gravitational collapse while the general-relativistic gravity gives a
destabilizing effect. The point to be emphasized is that the energy
for these two effects could have the same order of magnitude: In the
presence of rotation, the condition for the onset of the
general-relativistic instability is significantly different from the
well-known result for {\em spherical} stars derived by
Chandrasekhar~(1964). Indeed, fully general-relativistic study by
Baumgarte \& Shapiro~(1999) showed that rotation would be the
important ingredient, in their study for the stability of SMSs that
were modeled by a simple $\Gamma=4/3$ polytrope.

The purpose of this paper is to provide a quantitative formula for the
stability condition of rotating SMSs, which are supported by radiation
and gas pressure as well as by rotational centrifugal force.  We
assume that SMSs are rigidly rotating, because their cores in 
nuclear-burning phases should be in convective equilibrium (Bond et
al.,~1984; Umeda, private communication; see also the appendix of Loeb
\& Rasio,~1994) and hence they would be in a turbulent state.  We
systematically compute a number of equilibrium sequences for rotating
SMSs in general relativity employing polytropic equations of state
with its polytropic index, $n_p$, slightly smaller than 3 (i.e.,~the
adiabatic index slightly larger than 4/3), by which the equations of
state for SMSs are well-reproduced (see \S\,2).

The paper is organized as follows.  In \S\,2, we review approximate
equations of state for SMSs following Bond et al.~(1984).  In \S\,3,
the secular stability of rotating SMSs in general relativity is
numerically determined.  In \S\,4, we predict the final outcomes after 
the collapse of SMSs, assuming that the initial condition is a
marginally stable SMS determined in \S\,3.  Section~5 is devoted to a
summary and discussion.  Throughout this paper, $G$, $c$, $k_B$, and
$a_r$ denote the gravitational constant, speed of light, Boltzmann's
constant, and radiation constant, respectively.

\section{Equations of state}

We basically suppose that SMSs are composed of hydrogen, helium,
electron, and photon.  Then, the pressure, $P$, and internal energy
density, $\epsilon$, are written as (Bond et al.,~1984)
\beqn
P&=&{a_r T^4 \over 3} + Y_T n k_B T,\label{pressure}\\
\epsilon&=&a_r T^4 + {3 \over 2} Y_T n k_B T,\label{epsilon}
\eeqn
where $T$ is the temperature and $n$ is the baryon number density,
respectively. $Y_T$ is defined by
\beq
Y_T \equiv Y_e + Y_p + Y_\alpha,
\eeq
where $Y_I=n_I/n$ and $n_I$ for $I~=e, p, \alpha$ denotes the number
density of electron, hydrogen, and helium, respectively. For the
primordial gas, $Y_p \approx 0.75$, $X_\alpha\equiv 4Y_\alpha \approx
0.25$, and $Y_e \approx 0.88$, yielding $Y_T \approx 1.69$.  For pure
helium gas, $Y_p=0$, $X_\alpha=1$, $Y_e=0.50$ yielding $Y_T=0.75$.

Inside the SMSs in nuclear-burning phases, in particular for their
core region, convection should be highly enhanced, and a convective
equilibrium is realized (Bond et al.,~1984; Umeda, private
communication). This implies that the SMS core is isentropic, i.e.,
the specific entropy $s$ is constant, and its chemical composition is
uniform, i.e., $Y_T=$const.  For simplicity, we assume that these
relations are satisfied for the entire SMS, or we may say that we
focus only on the convective cores ignoring a surrounding
low-density envelope.

Then, the first law of thermodynamics, $d(\epsilon/n)=-Pd(1/n)$, gives
the relation between $dT$ and $dn$ (i.e., between $dP$ and $dn$) from
equations~(\ref{pressure}) and~(\ref{epsilon}).  Using this relation,
the adiabatic constant is calculated as (Eddington,~1918;
Chandrasekhar,~1939; Bond et al.,~1984)
\beq
\Gamma=\left({\pa \ln P \over \pa \ln n}\right)_{s}
={4 \over 3}+{4\sigma + 1 \over 3(\sigma+1)(8\sigma+1)},
\eeq
where $\sigma$ is the ratio of the radiation pressure to the gas
pressure, written as
\beq
\sigma \equiv {a_r T^3 \over 3Y_T n k_B}={s_\gamma \over 4Y_T k_B}.
\label{sigmadef}
\eeq
Here, $s_\gamma$ denotes the photon entropy per baryon.  For SMSs,
$\sigma$ and $s_\gamma/k_B$ are much larger than unity (see below),
and hence, $\Gamma$ can be approximated well by $4/3 + 1/(6\sigma)$.

Because the photon entropy is much larger than the gas entropy and $s$
is assumed to be constant, we may also assume that $s_\gamma$ and
$\sigma$ are approximately constant.  Hence, it is reasonable to
assume that the equations of state for the SMS core are well
approximated by a polytropic form
\beq
P=K\rho^\Gamma,~~~~\Gamma=1+{1 \over n_p}, \label{poly}
\eeq
where $\rho$ is the rest-mass density ($\rho=m_B n$ with $m_B$ the
mean baryon mass). $K$ and $n_p$ are the adiabatic constant and
polytropic index, respectively. 

Using equations~(\ref{pressure}) and~(\ref{poly}), the adiabatic 
constant is written as
\beq
K=\left({Y_T k_B \sigma \over m_B}\right)^{4/3}
\left({3 \over a_r}\right)^{1/3}
\left(1+\sigma^{-1}\right)\rho^{-1/(6\sigma)}. 
\eeq
Here, $\rho^{1/(6\sigma)}$ may be considered to be constant because
$\sigma \gg 1$ so that we can consider $K$ to be a constant.  From
$K$, the quantity of mass dimension is constructed as
\beq
M_u \equiv K^{n_p/2} G^{-3/2}c^{3-n_p}.\label{massu}
\eeq
For $\sigma \gg 1$, this quantity is written as
\beq
M_u=M_{u,3}\left(1+{3 \over 2\sigma} \right)
\left({m_B c^2 \over Y_T k_B T \sigma}\right)^{3/(4\sigma)},
\label{massunit}
\eeq
where $M_{u,3}$ denotes $M_u$ for $n_p=3$ ($\Gamma=4/3$) and is written as
\beqn
M_{u,3}&=&\left({Y_T k_B \over m_B}\right)^2
\left({3 \over G^3 a_r}\right)^{1/2} \sigma^2  \nonumber \\
&\approx& 4.01M_\odot Y_T^2 \sigma^2 \approx 0.251M_\odot
\left({s_\gamma \over k_B}\right)^2. \label{mass20} 
\eeqn
To derive equation~(\ref{massunit}), we used
equation~(\ref{sigmadef}).  Note that for $n_p=3$ polytropic spherical
stars in Newtonian gravity, the mass is written in the well-known form
as (Bond et al.,~1984)
\beq
M=C_3 M_{u,3}\approx 
1.14 M_\odot \left({s_\gamma \over k_B}\right)^2, \label{mass2}
\eeq
where we define $C_{n_p}\equiv M/M_u$ for each value of $n_p$, which 
will be determined in numerical analysis.  In the
present context, $C_{n_p}$ should be determined for SMSs at marginally
stable states. For the spherical case, it decreases slowly with the
decrease of $n_p$ from $C_3=4.555\equiv C_{3,s}$ to $C_{2.94} \approx 
3.66$ for the change from $n_p=3$ to $n_p=2.94$~($\Gamma \approx
1.340$) (see Table~1).

We note that the correction factor associated with $\sigma^{-1}$ in
equation~(\ref{massunit}) is a small number for SMS cores for which
$\sigma > 10^2$, although the typical value of $m_B c^2/(Y_T k_B T)$
is of order $10^4$: For example, for $Y_T=1.69$, $T=10^{8.2}$\,K, and
$\sigma \geq 10^2$, $1 \leq M_u/M_{u,3} \alt 1.06$: hence, $C_{n_p}
\approx M/M_{u,3}$. Here, we employ $T=10^{8.2}$\,K as the typical
temperature, supposing that the SMS is in a hydrogen-burning phase
(Bond et al.,~1984; Umeda, private communication) and assuming that
the hydrogen-burning temperature depends only weakly on its mass and
angular momentum.  If the SMS is in a helium-burning phase, $T$ should
be slightly higher as $\sim 10^{8.4}$\,K.

Using equation~(\ref{mass20}), we have
\beqn 
&&\Gamma-{4\over 3} \approx {1\over 6\sigma} \approx 3.8\times 10^{-3}
\left({M \over 10^5M_\odot}\right)^{-1/2} \nonumber
\\ && \hspace{3.cm} \times \left({C_{n_p} \over C_{3,s}}\right)^{1/2}
\left({Y_T \over 1.69}\right). \label{deltaG} 
\eeqn
This relation will be used in the next section.  We note that
$(C_{n_p}/C_{3,s})^{1/2}$ is in a narrow range between $0.90$ and $1.00$ for
$2.94 \leq n_p \leq 3$ (see Table~1).

Because we often refer to it later, we also analyze the magnitude of a
dimensionless quantity, $P/(\rho c^2)$. This is approximately 
written as
\beqn
{P \over \rho c^2} &\approx& {a T^4 \over 3\rho c^2} 
={1 \over 4}\left({k_B T \over m_B c^2}\right)
\left({s_\gamma \over k_B}\right) \nonumber \\
&\approx& 3.7 \times 10^{-6}\left({T \over 10^{8.2}\,{\rm K}}\right)
\left({s_\gamma \over k_B}\right). 
\eeqn
Using equation~(\ref{mass20}) with $\sigma \gg 1$, we find for
spherical SMS cores
\beqn
&&{P \over \rho c^2} \approx 1.1 \times 10^{-3}
\left({T \over 10^{8.2}\,{\rm K}}\right) \nonumber \\
&&\hspace{1.5cm} \times 
\left({C_{n_p} \over C_{3,s}}\right)^{-1/2}
\left({M \over 10^{5}M_\odot}\right)^{1/2}.
\label{pcrc}
\eeqn
For $n_p=3$ spherical polytropes in Newtonian gravity, the central
value of $P/\rho$, $P_c/\rho_c$, is equal to $(GM/R)/1.1705$ where $R$
is the stellar radius. Thus, the typical compactness of SMS cores 
defined by $GM/(c^2R)$ is of order $10^{-3}$ for $M \sim 10^5M_\odot$. 

\section{Numerical analysis for stability} 

\subsection{Basic equations}

To explore the secular stability of rotating SMSs against
general-relativistic quasi-radial collapse, we systematically compute
their equilibrium solutions in general relativity.  Assuming that the
SMSs are composed of an ideal fluid, we write the stress-energy tensor
as
\beqn
T^{\mu\nu}=\rho h u^{\mu} u^{\nu} + P g^{\mu\nu},
\eeqn
where $u^{\mu}$ is the four velocity, $h \equiv 1+ \varepsilon +
P/\rho$ is the specific enthalpy, $\varepsilon$ is the specific
internal energy (different from $\epsilon$), and $g_{\mu\nu}$ is the
spacetime metric.  As in \S\,2, the polytropic equations of state are
employed here.  Using the first law of thermodynamics, $\varepsilon$
in the polytropic equations of state is written as
\beq
\varepsilon={n_p P \over \rho}. 
\eeq
As mentioned in \S\,2, the SMSs (SMS cores) are likely to be in
convective equilibrium. This indicates that a turbulent state would be
realized and angular velocity would be approximately uniform
(Baumgarte \& Shapiro,~1999).  Thus, we pay attention only to rigidly
rotating stars setting the angular velocity $\Omega \equiv
u^{\varphi}/u^t$ to be constant.

With the polytropic equation of state, physical units enter the
problem only through the polytropic constant $K$, which can be
completely scaled out of the problem. For example, 
$K^{n_p/2}G^{-3/2}c^{3-n_p}(=M_u)$ has units of mass,
$K^{n_p}G^{-2}c^{5-2n_p}$ has units of angular momentum, and
$K^{-n_p}c^{2n_p}$ has units of density.  Thus, for presenting
numerical results, we show dimensionless quantities, which are
rescaled by $K$. We note that in these units, the dimensionless mass
$M$ (i.e.,~$M/M_u$) is equivalent to $C_{n_p}$, which is defined in
\S\,2.

Following Butterworth \& Ipser~(1976) (see also Stergioulas,~1998 for
a review), the line element is written as
\beqn
ds^2 &=& -e^{2\nu} dt^2
+ B^2 e^{-2\nu} r^2 \sin^2\theta (d\varphi - \omega dt)^2 \nonumber \\
&& +e^{2\zeta-2\nu}(dr^2 + r^2 d\theta^2), 
\eeqn
where $\nu$, $B$, $\omega$, and $\zeta$ are field functions. The first
three obey elliptic-type equations in axial symmetry, and the last one
an ordinary differential equation. These equations are solved using
two methods: one is described in Shibata \& Sasaki~(1998) and the
other is a method by Cook et al.~(1992). We checked that the results
derived by two independent codes agree well with each other for the
problems considered in this paper: For instance, the mass and density
of the turning points (see below) determined by two methods agree with
each other within 0.01\% and 1\% difference for most of the parameter
space (except for the region very close to the mass-shedding limit at
which it is not easy to identify the turning points).


The Komar mass (gravitational mass) $M$, Komar angular momentum $J$,
rotational kinetic energy $T_{\rm rot}$, and gravitational potential
energy $W$ are defined by
\beqn
M &=&2\pi \int (-2T_t^{~t}+T_{\mu}^{~\mu}) B e^{2\zeta-2\nu}
r^2 dr d(\cos\theta),~~\\ 
J  &=&2\pi \int \rho h u^t u_{\varphi} B e^{2\zeta-2\nu}
r^2 dr d(\cos\theta),\\ 
T_{\rm rot}&=&{1 \over 2} J \Omega,\\
W&=& M_{\rm p} - M + T_{\rm rot} ~(>0), 
\eeqn
where $M_p$ is the proper mass defined by
\beq
M_{\rm p} =2\pi \int \rho u^t (1+\varepsilon)
B e^{2\zeta-2\nu} r^2 dr d(\cos\theta). 
\eeq
Note that we define $W$ to be positive.  From these quantities, the
well-known dimensionless parameters are defined as $\beta \equiv
T_{\rm rot}/W$ and $q \equiv cJ/(GM^2)$.


In addition to these quantities, we often refer to the central
density, $\rho_c$, which is used to specify a rotating star for a
given set of $\beta$ and $M$, and to the equatorial circumferential
radius, $R_e$, by which a compactness parameter is defined by
$GM/(c^2R_e)$.  We also refer to a dimensionless quantity,
$P_c/(\rho_cc^2)$, for specifying the compactness of rotating SMSs. As
shown in the following, the values of this quantity are close to
$GM/(c^2R_e)$. 

\subsection{Analysis for secular stability}

The secular stability for rigidly rotating stars against quasi-radial
oscillations can be determined by the turning-point method established
by Friedman et al.~(1988) (see, e.g., Cook et al.,~1992;~1994;
Baumgarte \& Shapiro,~1999; Shibata,~2004 for application).  According
to the turning-point theorem, a change of the sign of $dM/d\rho_c$
along a curve of a constant value of $J$ indicates the change of the
secular stability.  Thus, in numerical computation, we derive curves
of constant values of $J$ in the plane composed of $M$ and $\rho_c$,
and then, determine the turning points. We always find one turning
point of $dM/d\rho_c=0$ along the $J=$const curves if $J$ is smaller
than a threshold value. In the present case, the lower-density side is
the branch for the stable stars and the other side is the branch for
the unstable stars. The stable branch has $(dM/d\rho_c)_J >0$, while
for the unstable branch, it is negative.

\subsection{Numerical results}

\begin{figure*}[t]
\begin{center}
\epsfxsize=3.4in
\leavevmode
\epsffile{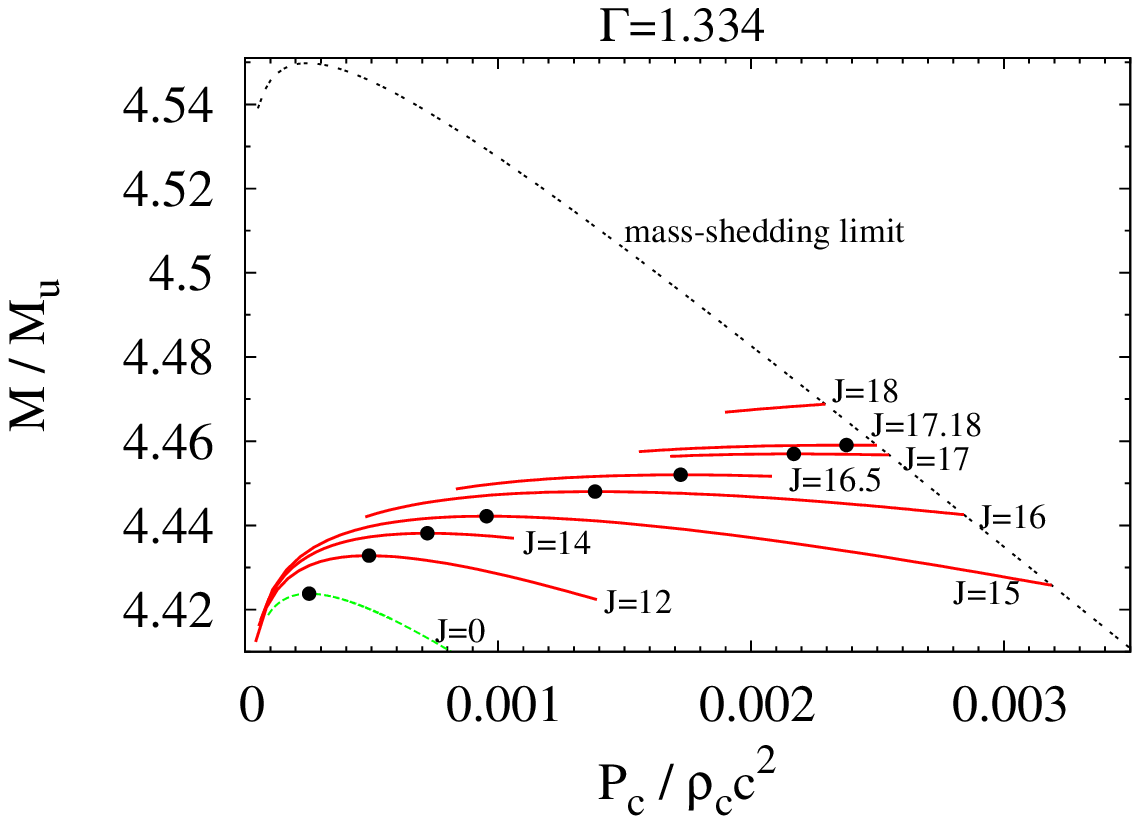}~~
\epsfxsize=3.4in
\leavevmode
\epsffile{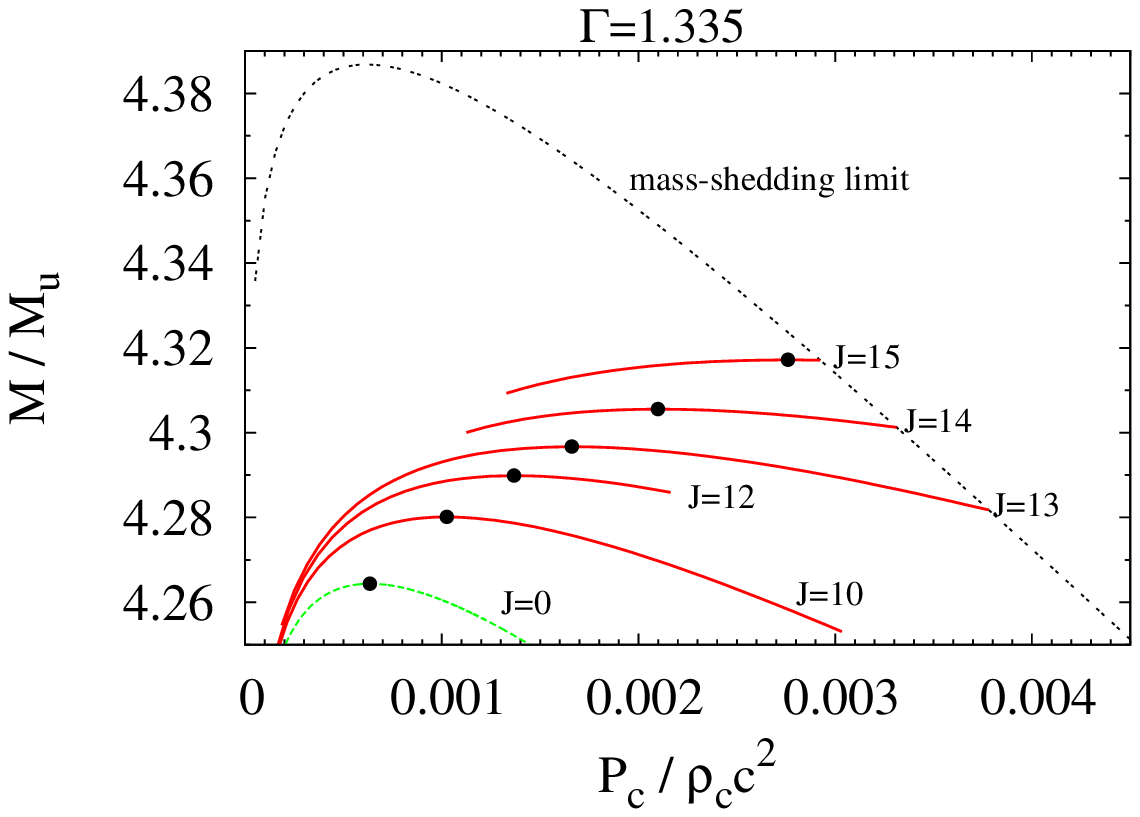}\\
\epsfxsize=3.4in
\leavevmode
\epsffile{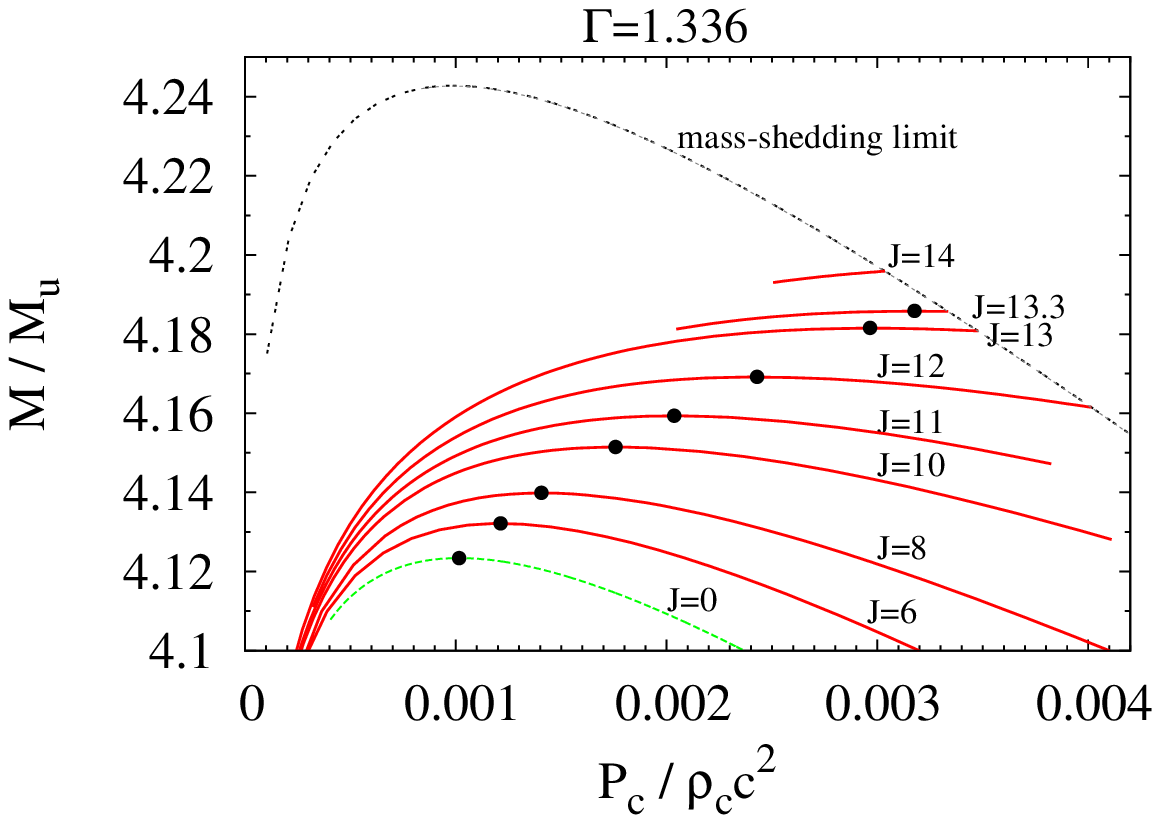}~~
\epsfxsize=3.4in
\leavevmode
\epsffile{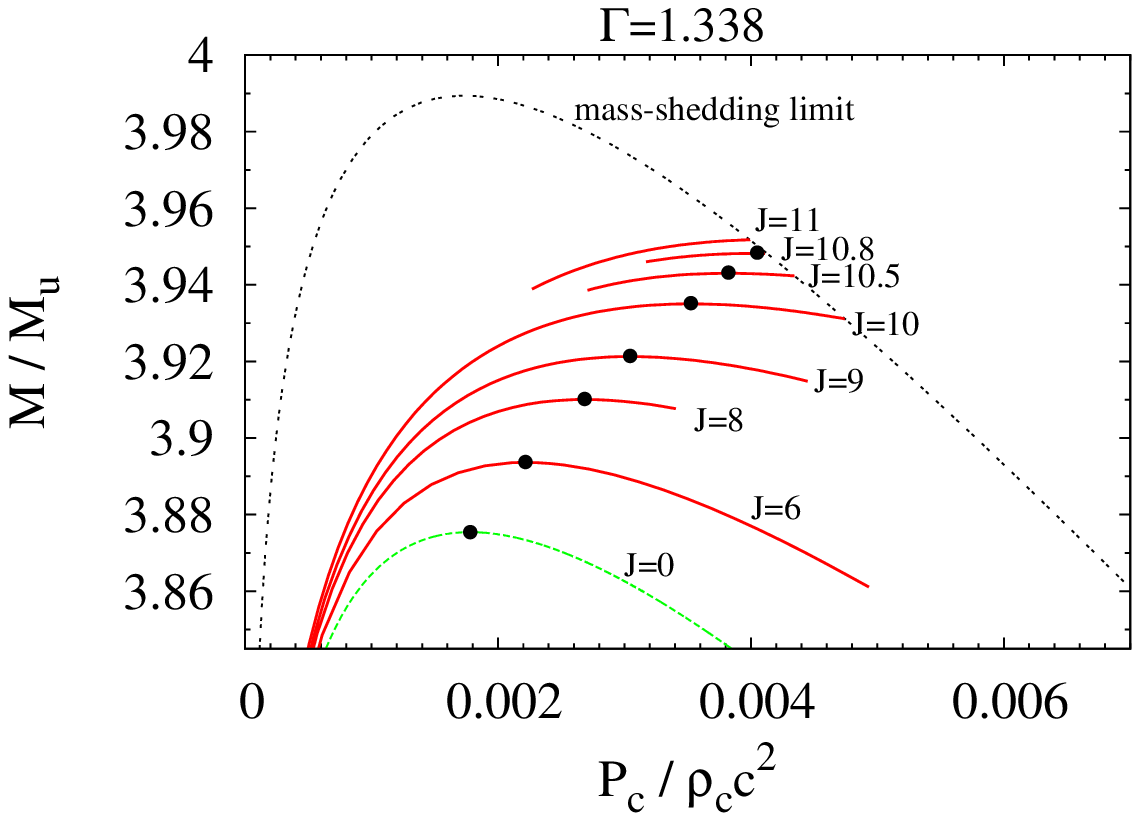}\\
\epsfxsize=3.4in
\leavevmode
\epsffile{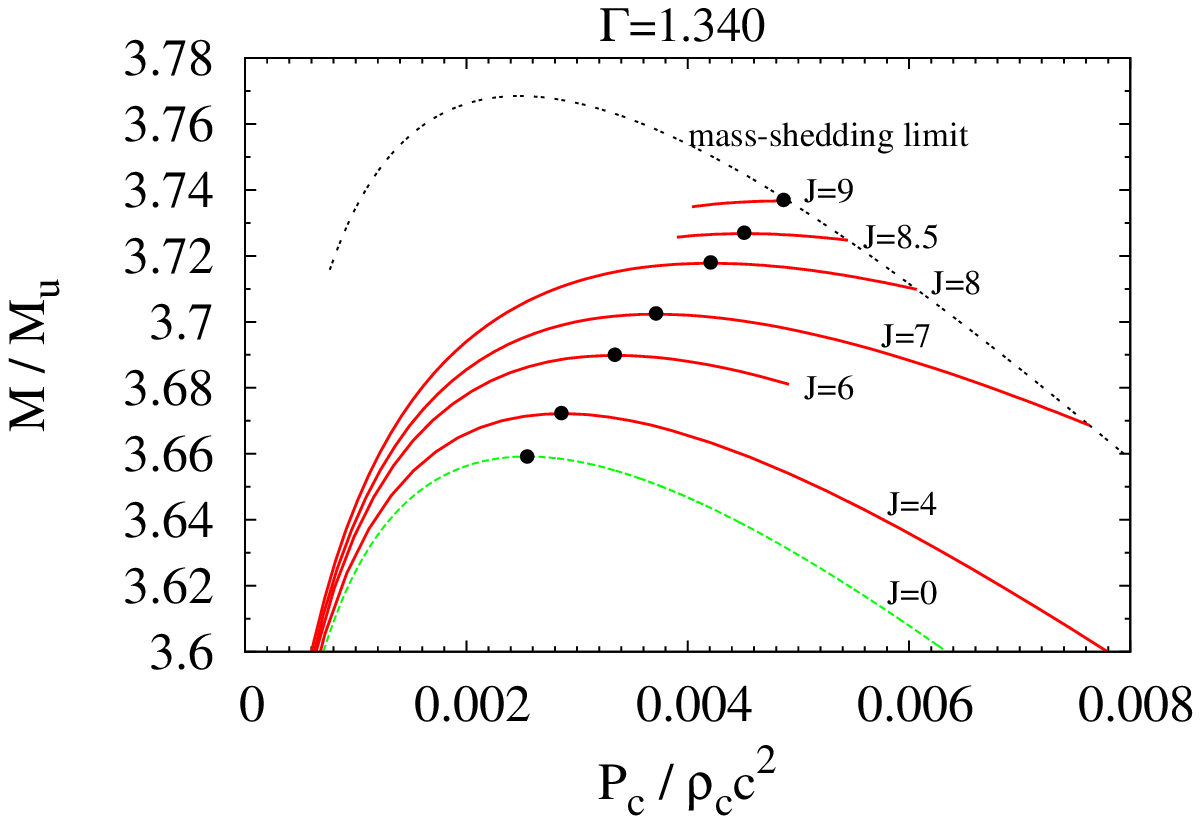}~~
\epsfxsize=3.4in
\leavevmode
\epsffile{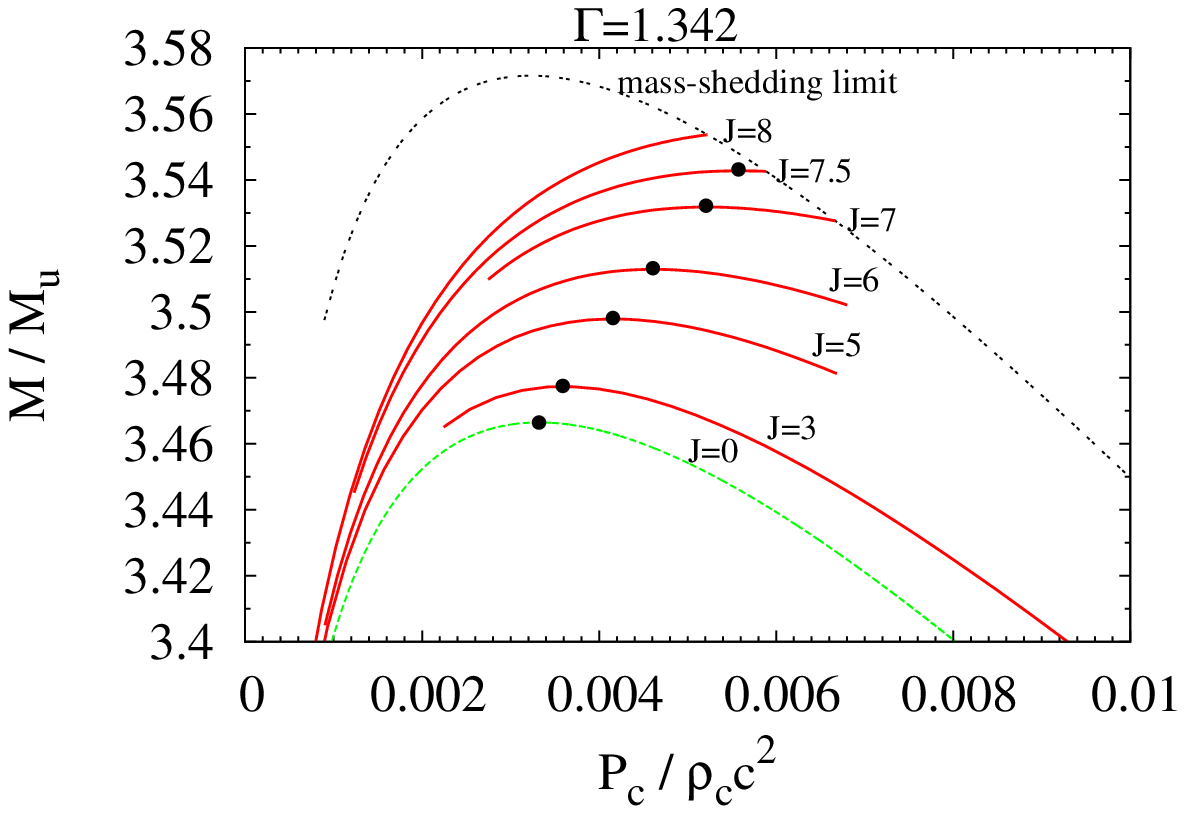}
\caption{Gravitational mass $M$ as a function of $P_c/(\rho_cc^2)$ for
  fixed values of $J$ (solid curves; and dotted curve for $J=0$) and
  for sequences of rotating stars at the mass-shedding limits (dot-dot
  curves) for $\Gamma=1.334$, 1.335, 1.336, 1.338, 1.340, and
  1.342. The units of the mass and angular momentum are
  $M_u=K^{n_p/2}G^{-3/2}c^{3-n_p}$ and $K^{n_p}G^{-2}c^{5-2n_p}$,
  respectively. The filled circles denote the turning points along the
  several curves of constant angular momentum.
\label{FIG1}
}
\end{center}
\end{figure*}

\begin{table*}[t]
\begin{center}
\caption{Maximum mass and related quantities for spherical SMSs and
  rigidly rotating SMSs.}
\begin{tabular}{ccccc}
\tableline\tableline
$\Gamma$ & $M_{\rm s,max}~(\hat C_{n_p})$ & $M_{\rm max}~(\hat C_{n_p})$ 
& $\beta_{\rm max}\,(\times 10^{-3})$ & $q_{\rm max}$  
\\ \tableline
1.334 & 4.424~(0.971) & 4.461~(0.979) & 8.9 & 0.87
\\ \tableline 
1.335 & 4.264~(0.936) & 4.319~(0.948) & 9.0 & 0.81
\\ \tableline
1.336 & 4.123~(0.905) & 4.188~(0.919) & 9.1 & 0.77
\\ \tableline
1.338 & 3.875~(0.851) & 3.950~(0.867) & 9.2 & 0.70
\\ \tableline
1.340 & 3.659~(0.803) & 3.737~(0.820) & 9.4 & 0.65
\\ \tableline 
1.342 & 3.466~(0.761) & 3.546~(0.778) & 9.6 & 0.61 \\
\tableline 
\end{tabular}
\tablecomments{$\Gamma$: adiabatic index.  $M_{\rm s,max}$: the
  maximum mass for spherical SMSs.  $M_{\rm max}$: the maximum mass
  for rigidly rotating SMSs. $\beta_{\rm max}$: the maximum value of
  $\beta$ for rigidly rotating SMSs. $q_{\rm max}$: the maximum value
  of $q$ for rigidly rotating SMSs. The units of $M_{\rm s,max}$ and
  $M_{\rm max}$ are $M_u=K^{n_p/2}G^{-3/2}c^{3-n_p}$ (i.e., we list
  $C_{n_p}$). Here, $\hat C_{n_p}=C_{n_p}/C_{3,s}$ with
  $C_{3,s}=4.555$. }
\end{center}
\end{table*}

\begin{figure*}[t]
\begin{center}
\epsfxsize=4.5in 
\leavevmode \epsffile{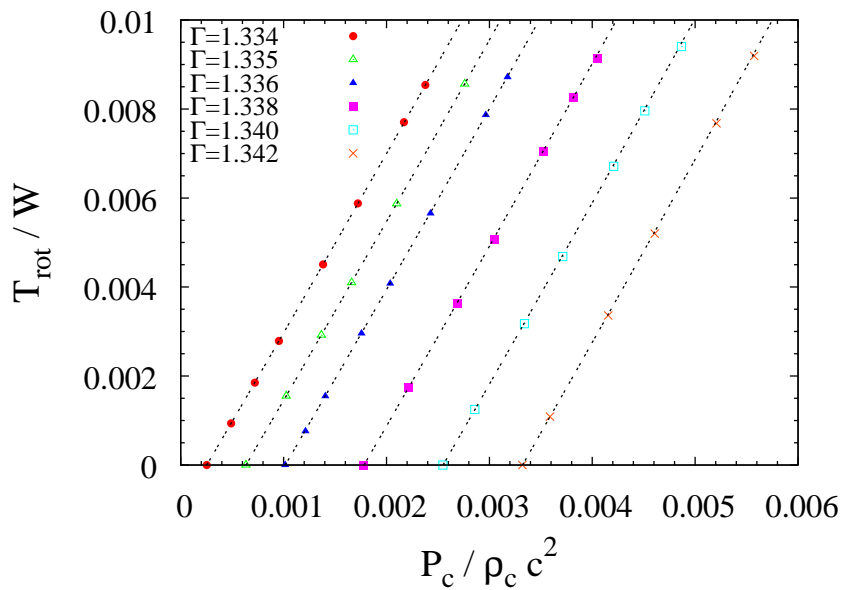}
\caption{The relation between $\beta(=T_{\rm rot}/W)$ and
  $P_c/(\rho_cc^2)$ for marginally stable stars with $\Gamma=1.334$,
  1.335, 1.336, 1.338, 1.340, and 1.342. Each dot-dot curve denotes the
  fitting formula of equation~(\ref{final}). For each value of
  $\Gamma$, SMSs located in the right-hand side of this curve are
  unstable to general-relativistic quasi-radial collapse.
\label{FIG2}
}
\end{center}
\end{figure*}

Figure~1 plots the curves of $M$ as a function of
$P_c/(\rho_cc^2)(=K\rho_c^{1/n_p}/c^2)$ for various values of $J$ and
for $\Gamma=1.334$, 1.335, 1.336, 1.338, 1.340, and 1.342.  Here, the
units of $M$ and $J$ are $M_u$ and $K^{n_p}G^{-2}c^{5-2n_p}$,
respectively.  For all the panels, the solid and dotted curves show
the relations of $M$ as a function of $P_c/(\rho_cc^2)$ for a given
value of $J$ and $J=0$, respectively, and the dot-dot curve denotes
the mass-shedding limit (i.e., the sequence of maximally and rigidly
rotating SMSs for a given equation of state): Above the dot-dot
curves, no rigidly rotating SMS can be realized.

The maxima of $M$ along $J=$const sequences are present for
$J/(K^{n_p}G^{-2}c^{5-2n_p}) \alt 17.3$, 15.2, 13.4, 10.9, 9.0, and
7.6 for $\Gamma=1.334$, 1.335, 1.336, 1.338, 1.340, and 1.342,
respectively.  If the value of $P_c/(\rho_cc^2)$ is smaller than that
at this turning point, the SMS is stable against general-relativistic
quasi-radial collapse.  On the other hand, if $P_c/(\rho_cc^2)$ is
larger than that at the turning point, the SMS is unstable: Rotating
SMSs will be unstable if they reach this turning point after some
evolution process of increasing the value of $P_c/(\rho_cc^2)$
(i.e.,~increasing the compactness or mass).  The values of the
dimensionless mass, $M/M_u$ ($C_{n_p}$), at the maxima depend only
weakly on the values of $J$ for a given value of $\Gamma$.  On the
other hand, the critical values of $P_c/(\rho_cc^2)$ depend strongly
on the value of $J$, in particular, for $\Gamma$ close to $4/3$. This
implies that for more rapidly rotating SMSs, the quasi-radial
instability sets in at a higher value of compactness. That is, to
induce the collapse of a rotating SMS, an additional evolution process
of increasing the compactness (this is equivalent to increasing the
mass; see below) is necessary.

Important quantities for marginally stable and maximally rotating SMSs
are listed in Table~1. Definitions of the quantities being tabulated
are provided in the {\small "NOTE"} that accompanies this table.  It
should be mentioned that the maximum values of $\beta$ for the stable
SMSs are universally $\sim 9 \times 10^{-3}$ depending weakly on
$\Gamma$ with its plausible values for SMSs.  By contrast, the maximum
values of $q$ depend strongly on the values of $\Gamma$, decreasing
far below the Kerr limit, $q=1$, with the increase of
$\Gamma$. $M_{\rm s,max}/M_u$ and $M_{\rm max}/M_u$ (i.e., $C_{n_p}$)
decrease slowly with $\Gamma$; they are well fitted in the form
$C_{n_p}=A-B(\Gamma-4/3)^\alpha$ where $(A, B, \alpha)\approx (4.60,
37.0, 0.734)$ for the spherical stars and $(A, B, \alpha)\approx
(4.60, 45.0, 0.789)$ for the maximally rotating stars, respectively.

\begin{figure*}[t]
\begin{center}
\epsfxsize=4.5in
\leavevmode
\epsffile{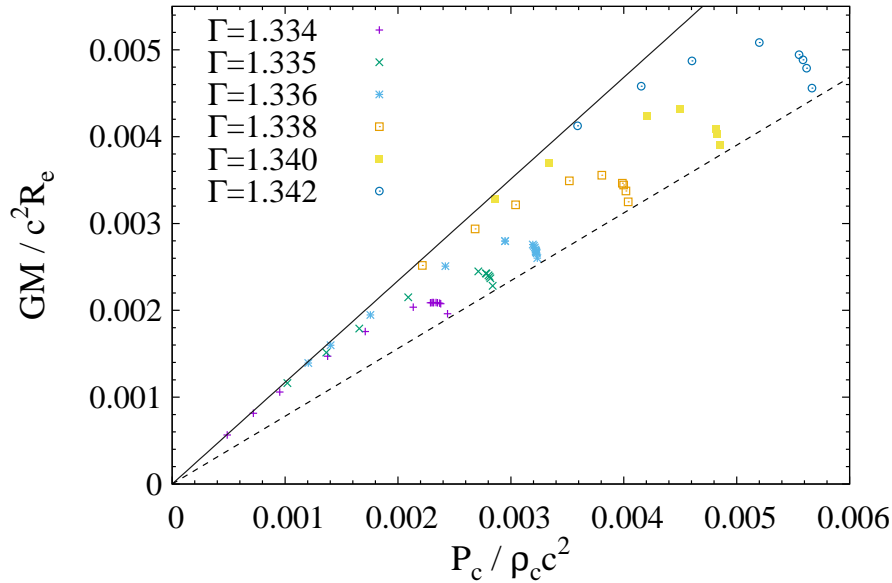}
\caption{The relation between $GM/(c^2R_e)$ and $P_c/(\rho_cc^2)$ for
  marginally stable SMSs. The solid line is
  $GM/(c^2R_e)=1.1705P_c/(\rho_cc^2)$, which is satisfied for $n_p=3$
  polytrope in Newtonian gravity. The dashed line shows the relation
  of $GM/(c^2R_e)=(2/3) \times 1.1705 P_c/(\rho_cc^2)$. For each curve
  of a fixed value of $\Gamma$, the values of $P_c/(\rho_c c^2)$
  increase with the increase of $\beta$ and the data approach the
  dashed line (go away from the solid line).
\label{FIG3}
}
\end{center}
\end{figure*}

Next, we derive a fitting formula for the stability condition of
rotating SMSs.  Figure~2 shows the relation between $\beta$ and
$P_c/(\rho_c c^2)$ for marginally stable SMSs: For each curve of
$\Gamma$, SMSs in its right-hand side are unstable.  It is found that
the relation is approximately linear, although it is slightly
different from the linear relation around the largest values of
$\beta$ (i.e.,~near the mass-shedding limit).

To construct a fitting formula, we first pay attention to the
spherical case, $\beta=0$. For $\Gamma \rightarrow 4/3$, Chandrasekhar
(1964) showed that the turning point would appear at
\beq
\Gamma-{4 \over 3}\approx 2.2489 {GM \over c^2R_e}.\label{zero}
\eeq
This is equivalent to 
\beq
\Gamma-{4 \over 3}\approx 2.6324 {P_c \over \rho_cc^2} \equiv y.
\eeq
However, for $\Gamma-4/3 \agt 10^{-3}$, the deviation from this
strictly linear relation is noticeable. By a high-precision numerical
analysis of spherical SMSs in general relativity, we find the
following better relation
\beq
\Gamma-{4 \over 3} + \left(\Gamma-{4 \over 3}\right)^2
\approx y, 
\eeq
or 
\beq
\Gamma-{4 \over 3} \approx y - y^2. 
\eeq

Then, we consider the case of $0 < \beta \ll 0.01$. Previous studies
in Newtonian gravity also have shown that the rotation plays a
significant role for the stabilization (Ledoux,~1945; see also
Tassoul,~1978 for a review) and the relation of equation~(\ref{zero})
is modified as
\beq
\Gamma-{4 \over 3}\approx y
-2\left({5 \over 3}-\Gamma\right)\beta.\label{first}
\eeq
Taking this condition into account and taking a close look at 
numerical results, we fit the numerical data 
of the turning points in terms of 
\beq
\Gamma-{4 \over 3}  
= y-y^2 - \left({10 \over 3}-2 \Gamma - y -\beta\right)\beta.\label{final}
\eeq
Here, we have determined the coefficients of the terms of $y\beta$ and
$\beta^2$ in a rather ad hoc manner: These coefficients could be
different from $-1$ in reality. However, this ad hoc choice is
acceptable in the present study because it provides a good fitting
formula as shown by the dot-dot curves in Figure~2 and this tells us
that the order of the magnitude of these coefficients is
unity. Indeed, the error estimated by
\beq 
{1 \over y}
\left[\Gamma-{4 \over 3} - y + y^2 + \left({10 \over 3}-2\Gamma -
  y - \beta\right) \beta\right]
\eeq
is always smaller than 1\% for our numerical data; in particular, for
the parameter space far from the mass-shedding limit, it is much
smaller than 1\%.  Therefore, we conclude that the fitting formula,
(\ref{final}), is well suited for determining the condition for the
onset of general-relativistic quasi-radial instability of 
rigidly rotating SMSs.

In equation~(\ref{final}), the coefficients of all the nonlinear
terms, $y^2$, $\beta^2$, and $y\beta$, are of order unity. This
implies that these nonlinear terms give only the minor effect on the
SMS stability because for (rigidly rotating) SMSs, $y \alt 0.01$ and
$\beta \alt 0.01$: We have therefore demonstrated that
equation~(\ref{first}), in essence the stability relation provided by
Tassoul (1978), can be used as an approximate condition.

Figure~3 plots the relation between $GM/(c^2R_e)$ and
$P_c/(\rho_cc^2)$ for SMSs at the turning points. This illustrates
that for $\beta \ll 0.01$ (i.e., for the limits of $GM/(c^2R_e)
\rightarrow 0$ and $P_c/(\rho_cc^2) \rightarrow 0$), the relation can
be approximated by
\beq
{GM \over c^2R_e}\approx 1.1705 {P_c \over \rho_cc^2}.
\eeq
As already mentioned, the factor $1.1705$ is derived from the analysis
of $n_p=3$ spherical polytropes in Newtonian gravity. For $\beta \agt
0.001$, the linear relation is not satisfied. For the limit that
$\beta$ approaches the maximum value $\beta_{\rm max}$ (i.e., for the
largest values of $P_c/(\rho_cc^2)$), the ratio of
$GM/(c^2 R_e)$ to $P_c/(\rho_c c^2)$ approaches $1.1705 \times 2/3$
(see the dashed line of Fig. 3). This stems from the fact that at the
mass-shedding limit, ratio of the polar axial length to the equatorial
circumferential radius is approximately $2/3$ depending very weakly on
the value of $n_p$.

Figure~3 shows that for $\Gamma=1.334$, the compactness of the SMSs at
the turning point increases by a factor of $\sim 7$ from the spherical
to rotating SMSs at the mass-shedding limit.  This factor decreases with
the increase of $\Gamma$. However, even for $\Gamma=1.335$--1.336
(these could be the typical values for a realistic SMS), this increase
factor is $\approx 2.3$--3.3. Therefore, the condition for the onset
of general-relativistic quasi-radial instability of rotating SMSs is
significantly different from that for spherical SMSs.  This fact has
to be taken into account for exploring the formation process of SMBHs
after the collapse of rotating SMSs.

Finally, we approximately determine the condition for the onset of
general-relativistic quasi-radial instability for realistic
SMSs. Specifically, the condition is imposed to the required minimum
mass for getting the instability.  Using equation~(\ref{pcrc}), $y$ is
written as
\beqn
y &\approx &2.9 \times 10^{-3}
\left({T \over 10^{8.2}\,{\rm K}}\right)
\left({C_{n_p} \over C_{3,s}}\right)^{-1/2}
\left({M \over 10^{5}M_\odot}\right)^{1/2} \nonumber \\
&=&A {\hat C}_{n_p}^{-1/2}M_5^{1/2}, \label{pcrc2}
\eeqn
where $M_5=M/10^5M_\odot$ and $\hat C_{n_p}=C_{n_p}/C_{3,s}$.
Substituting this equation and equation~(\ref{deltaG}) into
equation~(\ref{final}) and neglecting higher-order terms in $y$ and
$\beta$, we obtain the equation for $M_5$ as
\beqn
A {\hat C}_{n_p}^{-1} M_5 -{2 \over 3}\beta {\hat C}_{n_p}^{-1/2}M_5^{1/2} -B=0, 
\eeqn
where
\beqn
B=3.8\times 10^{-3} \left({Y_T \over 1.69}\right). 
\eeqn
Then, the solution for $M_5^{1/2}$ is
\beqn
M_5^{1/2}={\hat C_{n_p}^{1/2} \over 3A}\left(\beta+\sqrt{\beta^2 + 9AB}\right). 
\eeqn
If $M_5$ is larger than this value, SMSs are unstable.  Here, $\hat
C_{n_p}$ depends only weakly on $M$ and $\beta$ as already mentioned. 
Its plausible range is between 0.8 and 1.0; a more specific 
value can be obtained from equation~(\ref{deltaG}) and Table~1.

For $\beta=0$ with the plausible parameters for the SMS core in the
 hydrogen-burning phase, the threshold is
\beqn
M_5={B \over A} \approx 1.3
\left({T \over 10^{8.2}\,{\rm K}}\right)^{-1}
\left({Y_T \over 1.69}\right)\hat C_{n_p}. 
\eeqn
By contrast, for $\beta=0.009$ (i.e.,~near the mass-shedding limit),
$T=10^{8.2}$\,K and $Y_T=1.69$, $M_5 \approx 6.6\hat C_{n_p}$. Thus,
for obtaining unstable and maximally rotating SMSs, the mass has to be
increased by a factor of $\sim 5$ from that of the spherical SMSs at
the marginally stable point. Even for $\beta=0.005$, we obtain $M_5
\approx 3.4 \hat C_{n_p}$, and hence, significant mass increase is
necessary to get an unstable SMS. 

As touched on in \S\,1, SMSs could be formed in a high mass-accretion
environment. The often-referred highest mass-accretion rate is $\sim
0.1M_\odot$/yrs. The lifetime of the SMSs, which should shine
approximately at the Eddington limit, would be universally $\sim
2\times 10^6$\,yrs. This suggests that the typically final SMS mass
would be at most $\sim 2\times 10^5M_\odot$. Our present analysis
indicates that if they were appreciably rotating, the SMSs would not
collapse to a SMBH in their hydrogen-burning phase.

For a helium-burning phase and for $\beta=0$, $T \approx 10^{8.4}$\,K,
and $Y_T \approx 0.75$, we obtain $M_5 \approx 0.37\hat C_{n_p}$,
while for $\beta=0.009$, $T\approx 10^{8.4}$\,K, and $Y_T\approx 0.75$,
we obtain $M_5 \approx 2.4\hat C_{n_p}$. Thus, the mass of marginally
stable and maximally rotating SMSs has to be by a factor of $\sim 6$
larger than that of marginally stable spherical SMSs. For the
nonrotating case, general-relativistic quasi-radial collapse can be
induced even for $M_5 \sim 0.5$ (Chen et al.,~2014). However, for the
appreciably rotating case, it will not collapse by the general
relativistic instability in this phase.

For the oxygen-burning phase, the expected values are $T\approx
10^{9.0}$\,K and $Y_T\approx 0.56$.  Then, for $\beta=0$, $M_5 \approx
0.07\hat C_{n_p}$, and for $\beta=0.009$, we obtain $M_5 \approx
0.22\hat C_{n_p}$. Thus, for this case, a SMS core of relatively small
core mass can be unstable to general-relativistic quasi-radial
collapse.

\section{Predicting the final outcome}

\begin{figure*}[t]
\begin{center}
\epsfxsize=3.25in
\leavevmode
(a)\epsffile{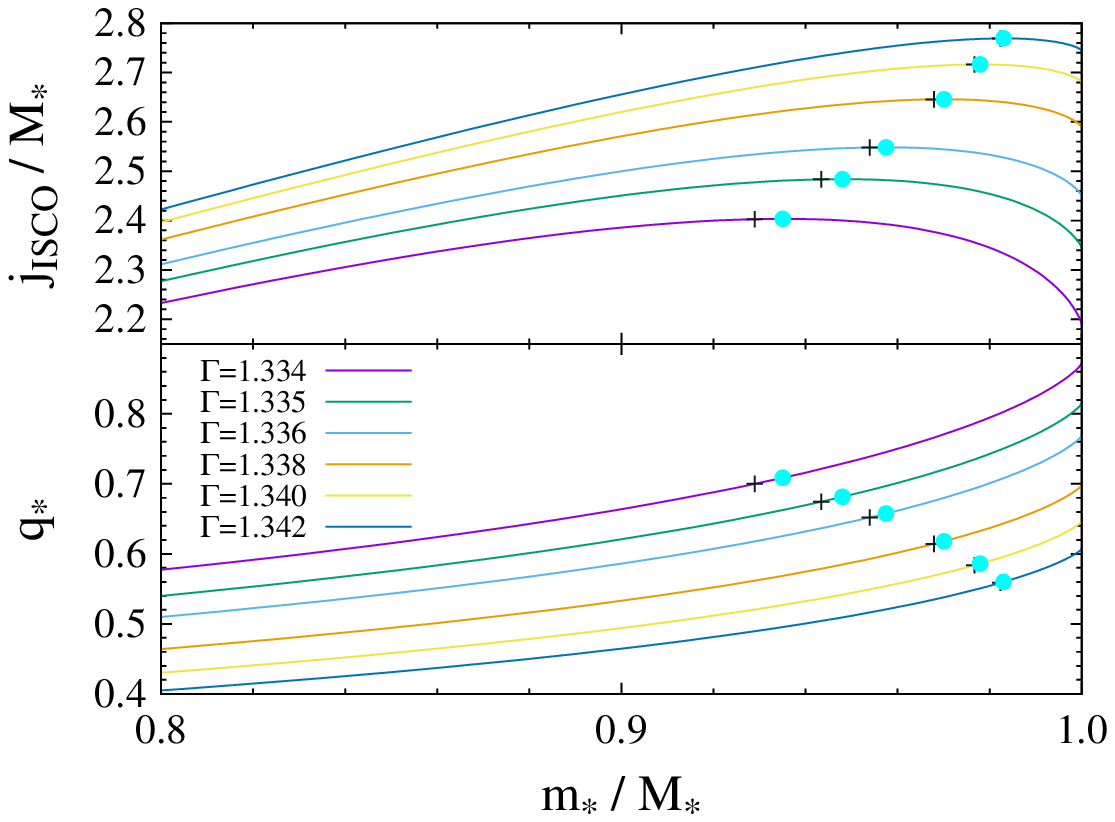}~~~~
\epsfxsize=3.25in
\leavevmode
(b)\epsffile{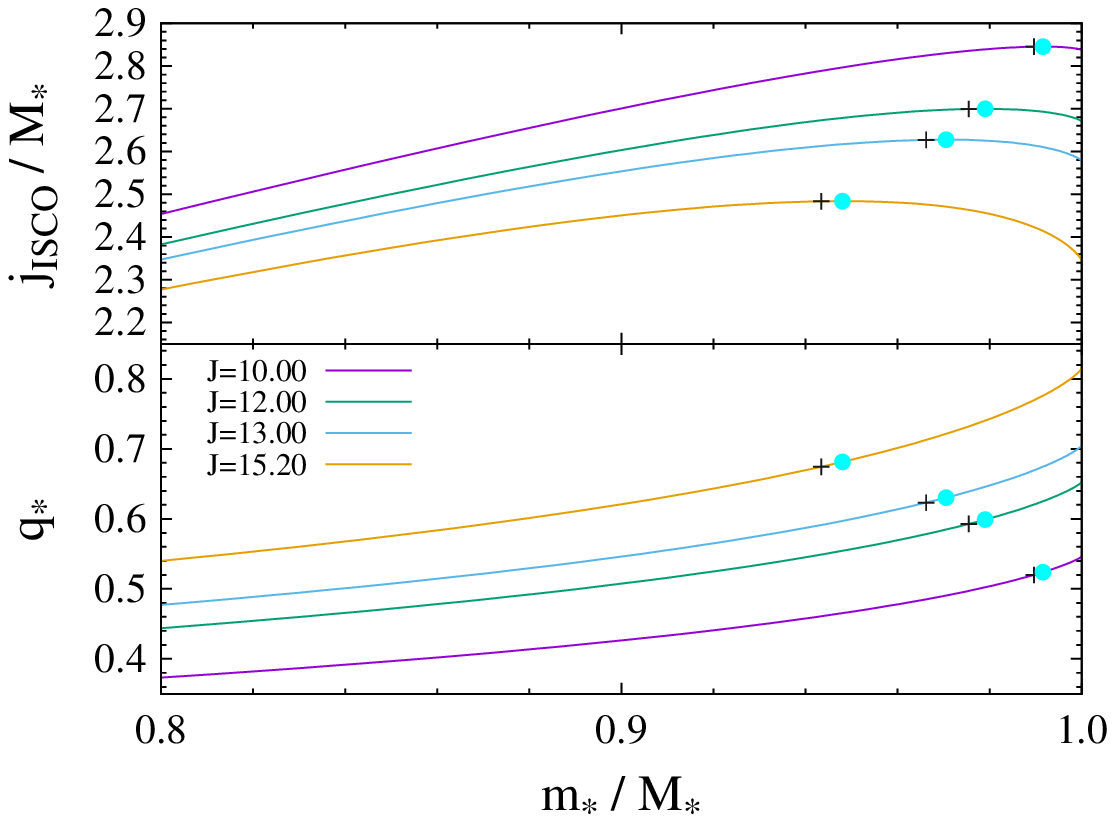}
\caption{(a) $j_{\rm ISCO}/m_*$ and $q_*$ as functions of $m_*/M_*$
  for marginally stable SMSs with $(\Gamma, J)=(1.334, 17.35)$,
  (1.335, 15.20), (1.336, 13.48), (1.338, 10.89), (1.340, 9.02), and
  (1.342, 7.63). Here the units of $J$ are
  $K^{n_p}G^{-2}c^{5-2n_p}$. The filled circles denote the maxima of
  $j_{\rm ISCO}/m_*$. The crosses denote the points at which $j=j_{\rm
    ISCO}$ is satisfied. (b) The same as the upper panel but for
  $\Gamma=1.335$ and $J=10$, 12, 13, and $15.2$ ($\beta \times
  10^3=1.54$, $2.92$, $4.09$, and $8.98$).
\label{FIG4}
}
\end{center}
\end{figure*}

The marginally stable SMSs determined in \S 3 are plausible initial
conditions for the collapse of SMS cores to a seed of a
SMBH. Following Shibata \& Shapiro (2002) and Shibata~(2004), we
predict the remnant of the collapse in the reasonable assumptions that
(i) the collapse proceeds in an axisymmetric manner, (ii) the viscous
angular momentum transport during the dynamical collapse is
negligible, and (iii) heating effects never halt the collapse.  The
numerical analysis is carried out in the same manner as that of
Shibata~(2004). We note that the assumption (iii) is not justified for
the case of an extremely high CNO abundance in a hydrogen-burning SMS
(Montero et al.,~2012) or for special SMS mass (Chen et al.,~2014).
We also note that in the presence of steeply differential rotation,
the centrifugal force could halt the collapse (Reisswig et al.,~2013),
although for the rigidly rotating case, we would not have to consider
this possibility (Shibata \& Shapiro,~2002; Liu et al.,~2007; Montero
et al.,~2012).

Since viscosity is assumed to be negligible during the collapse, 
the specific angular momentum $j$ of each fluid element is conserved
in axisymmetric systems. Here, $j$ is defined by 
\beq
j \equiv h u_{\varphi}, 
\eeq
and it increases with the increase of the cylindrical radius 
for the SMSs of $\Omega=$const. 

Next, we define rest-mass distribution as a function of $j$, $m_*(j)$, as
\beq
m_*(j_0) \equiv 2\pi \int_{j < j_0} \rho u^t B e^{2\zeta-2\nu} 
r^2 dr d(\cos\theta). 
\eeq
Here, the integration is performed only for the elements with $j < j_0$
for a given value of $j_0$.  In addition, we define the total angular
momentum with the specific angular momentum less than a given value
$j_0$:
\beqn
J(j_0)=2\pi \int_{j < j_0} \rho h u^t u_{\varphi} B e^{2\zeta-2\nu}
r^2 dr d(\cos\theta). 
\eeqn

Then, we assume that a seed black hole is formed during the collapse
and it dynamically grows with the subsequent infall of ambient
matter. For its growth process, we consider an innermost stable
circular orbit (ISCO) in the equatorial plane around the growing black
hole at the center. We then assume that the formed black hole grows
sequentially capturing fluid elements from lower values of $j$.  It is
natural to consider that if $j$ of a fluid element is smaller than the
value at this ISCO $(j_{\rm ISCO})$ of an instantaneous black hole,
the element will fall into the black hole eventually (as long as
$j_{\rm ISCO}$ increases with the black-hole growth).

To determine $j_{\rm ISCO}$, we assume that at each moment of the
black-hole growth, the instantaneous mass and angular momentum are
approximated by $m_*(j)$ and $J(j)$ with its dimensionless spin
parameter $q_*(j) \equiv J(j)/m_*(j)^2$. Note that the baryon rest
mass of SMSs is nearly equal to the gravitational mass because of
their soft equations of state with $n_p \approx 3$ and weak general
relativistic correction.  If we further assume that the spacetime can
be approximated instantaneously by a Kerr metric, we can compute
$j_{\rm ISCO}$ (Bardeen et al.,~1972; chapter 12 of Shapiro \&
Teukolsky,~1983).  The value of $j_{\rm ISCO}$ changes as the black
hole grows.  If $j_{\rm ISCO}$ increases, additional mass may fall
into the black hole. However, if $j_{\rm ISCO}$ decreases, ambient
fluid that has $j > j_{\rm ISCO}$ will no longer be captured. This
expectation suggests that when $j_{\rm ISCO}$ reaches a maximum value,
the dynamical growth of the black hole should have already
terminated. Thus, by this consideration, we can determine the possible
maximum mass of the black hole.  In reality, the growth of the black
hole may be terminated before reaching the maximum of $j_{\rm ISCO}$:
Because the mass infall is possible only for the case of $j < j_{\rm
  ISCO}$ at each instantaneous time, the mass accretion would
terminate if the point for $j=j_{\rm ISCO}$ is reached. In the
following, we consider these two possibilities as in Shibata (2004).



To illustrate the models for the growth of the black-hole mass and
dimensionless spin, in Figure 4, we plot $j_{\rm ISCO}[m_*,q_*]/M_*$
and $q_*$ (a) for $(\Gamma, J)=(1.334, 17.35)$, $(1.335, 15.20)$,
$(1.336, 13.48)$, $(1.338, 10.89)$, $(1.340, 9.02)$, and $(1.342,
7.63)$ and (b) for $\Gamma=1.335$ and $J=10$, 12, 13, and 15.2 ($\beta
\times 10^3=1.54$, $2.92$, $4.09$, and $8.98$).  Here, $M_* (\approx
M)$ denotes the total rest mass of the SMSs.  We choose the SMSs close
to marginally stable states at the mass-shedding limit for (a) while for
(b), the degree of the rotation is chosen for a wide range.  For the
models shown in Figure 4(a), the maximum of $j_{\rm ISCO}$ is reached
at $m_{*}/M_* \approx 0.935$, 0.948, 0.957, 0.970, 0.978, and 0.983
for $\Gamma=1.334$, 1.335, 1.336, 1.338, 1.340, and 1.342,
respectively ({\it circles} of Figure~4) while the condition of
$j=j_{\rm ISCO}$ is satisfied at $m_{*}/M_* \approx 0.929$, 0.943,
0.954, 0.968, 0.977, and 0.982 for $\Gamma=1.334$, 1.335, 1.336,
1.338, 1.340, and 1.342 respectively ({\it crosses} of Figure~4).
After the maximum of $j_{\rm ISCO}/M_*$ is reached, $j_{\rm ISCO}/M_*$
steeply decreases.  Thus, once the black hole reaches this point,
it will stop entirely growing dynamically. The dynamical growth may be
stopped when the point of $j=j_{\rm ISCO}$ is reached as already
mentioned. However, $m_*$ for this point is only slightly smaller than
that at the maximum of $j_{\rm ISCO}$.  This suggests that the
dynamical growth will be stopped near the maximum of $j_{\rm
  ISCO}$. In any case, more than 90\% (up to $\approx 98$\%) of the
SMS matter will fall into a SMBH dynamically (i.e., in a time scale
much shorter than dissipation and angular-momentum transport time
scales).  Lower panels of Figure 4(a) show that $q_* \approx 0.709$,
0.681, 0.658, 0.618, 0.586, and 0.560 at the maximum of $j_{\rm ISCO}$
and $q_* \approx 0.700$, 0.675, 0.652, 0.614, 0.584, and 0.559 at
$j=j_{\rm ISCO}$ for $\Gamma=1.334$, 1.335, 1.336, 1.338, 1.340, and
1.342, respectively.  This suggests that the SMBHs formed after the
dynamical collapse will not be rapidly rotating. 

Figure~4(b) shows that with the decrease of $J$ (and $\beta$), the
value of $m_*$ at the maximum of $j_{\rm ISCO}$ approaches $M_*$.
However, even for $\beta=O(10^{-3})$, $m_*$ is $\approx 0.99M_*$.
This property holds irrespective of $\Gamma$ with its plausible values
for SMSs.  The dimensionless spin of the formed black hole also
decreases with the decrease of $J$, although the remnant black hole is
still likely to be rotating with moderate spin even for the models
with $\beta=O(10^{-3})$.

Because a fraction of SMS matter does not fall directly into the SMBH,
after the dynamical collapse, a system of a black hole surrounded by
disks or tori in a dynamical state likely will be formed.  Subsequent
evolution of this system will be determined primarily by viscous
effects in the accretion disks/tori.  The analysis here suggests that
disk/torus mass is less than 10\% of the initial SMS mass even for
maximally rotating SMS models, and for $\beta=O(10^{-3})$, it is $\sim
1\%$.  However, this does not mean that the effect of the disk/torus
would be minor, because their mass could be $10^3$--$10^5M_\odot$ due
to the fact that the mass of the progenitor SMS core would be quite
large as $10^4$--$10^6M_\odot$.  Exploring the possible signals from
such a high-mass disk/torus by a numerical-relativity simulation is an
interesting subject in the future (see Liu et al.,~2007 for a previous
effort).

\section{Summary and discussion} 

We studied the secular stability of rigidly rotating SMSs against
quasi-radial collapse in general relativity. We showed that the
stability condition of SMSs depends appreciably on the ratio of the
rotational kinetic energy to the gravitational potential energy
(denoted by $\beta$).

Our result implies that for the onset of general-relativistic
quasi-radial collapse of rapidly rotating SMSs in the hydrogen-burning
phase, high SMS mass is necessary.  In particular, for the possible
maximum value of $\beta \sim 0.009$, the mass required for the
instability is by a factor of $\agt 5$ larger than that for the
spherical SMSs. Since the SMSs shine approximately at the Eddington
limit and their lifetime would be universally $\sim 2\times 10^6$\,yrs
irrespective of their mass, for producing such a high-mass rotating
SMS, a quite high mass-accretion rate would be necessary: e.g., for a
SMS of mass $6\times 10^5M_\odot$, the accretion rate has to be at
least $0.3M_\odot$/yrs; in reality, a much higher accretion rate would
be necessary because all the accreted matter does not form the SMS core.
This suggests that rapidly rotating SMSs that collapse to a SMBH via
general-relativistic quasi-radial instability would be rarer than
nonrotating or slowly rotating SMSs: Although rapidly rotating SMSs
could be formed, few of them would collapse to a SMBH via general
relativistic quasi-radial instability in the hydrogen-burning
phase. Rather, most of them will evolve as a result of hydrogen and
helium-burning, and eventually, the formed oxygen core could collapse
to a SMBH via the general-relativistic instability or pair instability
as in the less massive stars (Bond et al.,~1984).

If a large fraction of SMSs are rapidly rotating, the mass of a seed
of SMBHs thus formed from an oxygen core could be much smaller than
$10^5M_\odot$, and the seed SMBHs would be initially surrounded by a
huge amount of matter (in the absence of significant mass loss during
the nuclear-burning phases). A fraction of the ambient matter will
subsequently accrete onto the central SMBH and surrounding
disk/torus. Exploring this infall and growth phase of the SMBH is an
interesting subject for the future. In particular, exploring the
resulting signals is an important subject, because they could bring
information for the SMBH formation process (e.g., Matsumoto et
al.,~2015).

\acknowledgments

We are grateful to Hideyuki Umeda for showing us his preliminary
numerical results for the evolution of spherical SMSs in
nuclear-burning phases and for a helpful discussion.  We also thank
Takashi Yoshida for a fruitful conservation and our referee for
helpful comments.  This work was supported by Grant-in-Aid for
Scientific Research (24244028) of Japanese MEXT/JSPS.


\end{document}